\documentclass[12pt,reqno]{article}
\usepackage{amsmath,amsthm}

\input{tcilatex}

\begin{document}

\begin{center}
\textbf{\large Algebraic Quarks from the Tangent Bundle: Methodology\bigskip 
}

\textbf{{\large Jos\'{e} G. Vargas}}

PST Associates, LLC (USA)

{138 Promontory Rd., Columbia, SC 29209}\linebreak {josegvargas@earthlink.net%
}

To Bernd Schmeikal, for his pioneering work\linebreak on $SU(3)$ as
spacetime symmetry{\small \bigskip }
\end{center}

\textbf{Abstract.} In a previous paper, we developed a table of components
of algebraic solutions of a system of equations generated by an
inhomogeneous proper-value equation involving K\"{a}hler's total angular
momentum. This table looks as if it were a representation of real life
quarks. We did not consider all options for solutions of the system of
equations that gave rise to it. We shall not, therefore, claim that the
present distribution of those components as a well ordered table has strict
physical relevance. It, however, is of great interest for the purpose of
developing methodology, which may then be used for other solutions.

We insert into our present table concepts that parallel those of the
phenomenology of high energy physics (HEP): generations, color, flavor,
isospin, etc. Breaking then loose from that distribution, we consider
simpler alternatives for algebraic ``quarks'' of primary color (The
mathematics speaks of each generation having its own primary color). We use
them to show how stoichiometric argument allows one to reach what appear to
be esthetically appealing idempotent representation of particles for other
than electrons and positrons (K\"{a}hler already provided these half a
century ago with idempotents similar to our hypothetical quarks). We then
use neutron decay to obtain formulas for also hypothetical algebraic
neutrinos and the $Z_{0}$, and use pair annihilation to obtain formulas for
gamma particles.

We finally go back to the aforementioned system of equations and start to
develop an alternative option. We solve the system of equations for this new
option but stop short of studying it along the lines of the present paper.
This would thus be an easy entry point to this theory by HEP\ physicists.
Their knowledge of the phenomenology will allow them to go faster and
further.

\section{Introduction}

The present paper becomes the fourth in a series \cite{V51}, \cite{V55}, %
\cite{V56} on idempotents that enter solutions with symmetry of exterior
systems. This was first considered by \'{E}. K\"{a}hler \cite{Kahler61}, but
only for only for idempotents involving only scalar-valued differential
forms. We have studied Clifford-valued ones \cite{V55}, \cite{V56}. This
permits a formidable enrichment in the theory of such solutions.

In \cite{V51}, we reproduced a little geometric calculation by \'{E}. Cartan %
\cite{Cartan22}. It shows that modern differential geometry amounts to 
\textit{only }a theory of moving frames. We solved this problem through a
canonical Kaluza-Klein space (KK) where the fifth dimension is propertime.
No compactification is involved. Readers who see it difficult to conceive of
propertime as a fifth dimension need only accept that a fifth coordinate
becomes propertime on curves, where all non-null differential forms are
multiples of just one differential 1-form.

Time and propertime remain connected here. The interaction of these two
concepts takes place (a) in the product of two time-space-propertime
Clifford algebra structures (resulting in Clifford-valued K\"{a}hler
differential forms in five dimensions), followed by (b) algebra of
privileged elements in that product structure and (c) finally, idempotents
in space-propertime subspace.

The orthogonality of time-space frames that characterizes Special Relativity
(SR) ---and more generally the Lorentz transformations (LTs)---\ is not
present in its associated space-propertime, which is to be viewed as the
high energy physics (HEP) subspace. As argued in \cite{V51}, there is a
specific alternative to SR that preserves orthogonality in space-propertime
at a most fundamental level, and still involves the LTs at a practical
level. The role of these transformations in what might in principle be a
non-relativistic context but with length contraction and time dilation has
been known to philosophers of science and interested physicist for many
decades, in the context of the issue of conventionality of synchronizations.
Theirs has been such an extreme view that they have actually claimed that
there is no new theory in that alternative, a statement which goes too far.
The LTs are inescapable in this ``para-Lorentzian'' (PL) alternative, but
the SR and PL bundles of frames are different, which is of the essence in
our argument. Unexpectedly, the space-propertime of that alternative is like
the time-space of SR, thus orthogonal. This is not the case when the bundle
of spacetime frames is the set of orthonormal frames associated with the LTs.

Because of that structural equivalence of relativistic time-space and PL
space-propertime and in order to postpone anything controversial, we
developed mathematical theory in \cite{V55} and \cite{V56} as if\ we were
dealing with the time-space of SR [We, however, dealt extensively with the
physical differences in our paper ``$U(1)\times SU(2)$ from the tangent
bundle'' \cite{V51}]. We connect here with those papers by first giving
names (quarks, flavor, color, generations) to concepts that had previously
emerged \cite{V56}. This name-giving demands that one assigns specific
directions in space to the generations, since there has to be a reason why
the masses of ``alike particles'' differ as a function of generation. But
consideration of this issue far exceeds the reach of this paper. Suffice to
say for illustration that we have in mind the likes of the direction of the
velocity with respect to the preferred frame of the center of mass of a
system of colliding particles .

The contents of the paper is organized as follows. In section 2, we proceed
to discuss the limitations of group theory in physics, with the
retrospective view emerging in our study. We similarly ignore that the
connections that enter Yang-Mills theory live in auxiliary bundles not
directly related to the time-space manifold. Consider $U(1)$ on its own. It
does not represent all 1-dimensional groups, but $u(1)$ does. Hence there is
no problem in thinking of it as pertaining to time translations. Even less
of a problem lies in thinking of $SU(2)$ and, therefore, also of $su(2)$, as
pertaining to rotations. But $SU(3)$, or $su(3)$ for that matter, is not
directly connected with geometric motions like, say, displacements. If we,
however, look at \textit{algebraic structure }in the manifold of solutions,%
\textit{\ }we find idempotents that look like an algebraic representation of
quarks related to symmetry under translations and rotations \cite{V55}.
Hence, though groups have certainly been very helpful in theoretical
physics, they may not be the best road to structure in high energy physics
(HEP). The best venue for HEP\ may lie in algebra of solutions of equations
rather than in group theory approach to the symmetry of those equations (See
section 2). If we discount antiparticles, which K\"{a}hler showed to be a
concomitant of time translation symmetry \cite{Kahler62}, the 3-D Euclidean
group of symmetries suffices to get us to an algebraic palette of quarks.

Apart from the work of K\"{a}hler on idempotent solutions of his ``K\"{a}%
hler-Dirac equation'' \cite{Kahler61} ---to which he referred simply as
Dirac equation in spite of its paradigm changing character--- a major
inspiration for our work is a paper on algebraic quarks by Schmeikal \cite%
{Schmeikal}. The part of his work that occupies us in subsection 3.1 is of
algebraic nature. But our route to a (different) algebraic representation of
quarks is very different from his, as he uses only tangent Clifford algebra.
In subsection 3.2, we summarize in simple terms the nature of the structure
in which we expand K\"{a}hler's use of idempotents in solutions with
symmetry of equations.

In section 4, the algebraic results with which we concluded paper \cite{V56}
are connected with phenomenology. Color and flavor become algebraic concepts
directly tied to geometry. In section 5, we show how to combine
stoichiometry with particle reaction phenomenology. In doing so, we
implicitly make the case that quark stoichiometry may have a great potential
for improving present fundamental particle theory, in addition to connecting
it to tangent bundle geometry. In section 6, we study solutions with proper
value different from zero of the algebraic system that gave rise to the
palette of quarks that has occupied us in this paper.

To conclude this introduction, let us bring to the attention of conservative
readers that a new perspective on the LTs and concomitant new perspective on
flat time-space structure may be a very low price to pay in order to make
algebraic progress in high energy-physics (HEP). Hypothetical deviations
from the LTs at ultra-relativistic speeds, which is sometimes intimated in
the HEP literature, may not be necessary at all for theoretical progress in
certain areas of HEP. One simply may need to accept that the LTs are the
facade of something deeper. That is no problem; there is no great building
without a facade.

\section{Precursor Issues}

\subsection{Clifford-algebra related issues}

\subsubsection{Background field: a main difference between Dirac and K\"{a}%
hler}

Where Dirac uses a tangent structure to deal with quantum physics with
electromagnetic coupling, K\"{a}hler uses an algebra of scalar-valued
differential forms. We shall refer to the latter as K\"{a}hler algebra. This
leaves the tangent Clifford algebra for use as valuedness algebra. Unlike
the Dirac equation, which involves spinors, K\"{a}hler's equation is in
first instance about a ``background field''. Spinors are solutions with
symmetry emerging in that background field, which then is to be considered
as primordial. Probability densities must be considered as a derived
concept. This is compatible in principle with Einstein's view of particles
as regions of high concentration of the field. But it was naive of Einstein
to think that the type of equations he considered would yield particles.

Related to quantum equations being about a non-spinorial field, operators
are not in general as fundamental in K\"{a}hler's scheme. Except those for
energy and angular momentum, whose relevance is already known from group
theoretic arguments, operators also should be considered as emerging
concepts. As an example, consider K\"{a}hler's equation with electromagnetic
coupling \cite{Kahler61}, \cite{Kahler62}. Momenta and generalized momenta
operators emerge while obtaining a classical Hamiltonian with
electromagnetic coupling by the expedient process of considering the mass of
the electron as the dominant energy, which is introduced in the exponent of
a phase factor \cite{V44}.

Concepts like strangeness, charm, etc. are results of the inventiveness of
physicists. They have been created to represent reality, but there is always
room for better representations. In what concerns our algebraic
representation, our idempotents speak of where the field wants to go but
does not quite get because the particles that the ternary idempotents would
embody (read quarks) start their collapse in the same process as they
building their defining properties.

In K\"{a}hler's calculus (KC), idempotents become the principal factor in
the products that constitute solutions with symmetry of equations (See Eq.
11). Those solutions are composed of pieces, the set of which constitutes a
highly structured palette, unlike anything that the solutions of the Dirac
theory have to offer.

In HEP\ based on Dirac's theory, the structure of the set of quarks is
reached through phenomenology. In contrast, the ``primordial field'' of K%
\"{a}hler's quantum mechanics can materialize into a rich variety of linear
combinations of spacetime related idempotents, making superfluous the
resorting to internal symmetries and auxiliary bundles. Given the K\"{a}hler
ascendancy of those idempotents, they should be taken seriously.

\subsubsection{Idempotents in the commutative algebra arising from
geometrization of units imaginary}

K\"{a}hler considered the Clifford algebra of scalar-valued differential
forms based on the equations 
\begin{subequations}
\begin{equation}
dx^{\mu }dx^{\nu }+dx^{\nu }dx^{\mu }=2\eta ^{\nu \mu }
\end{equation}%
for the Lorentzian signature. Again, we refer to it as K\"{a}hler algebra,
regardless of signature and dimension. He also considered generalizations
where the differential forms are tensor-valued. He produced several physical
applications for scalar-valued differential forms, but none for more general
valuedness.

Clifford-valued clifforms are the elements of the product of two Clifford
algebras. One of them is K\"{a}hler's algebra. The other one is its
associated tangent Clifford algebra, defined by%
\begin{equation}
\mathbf{a}_{\mu }\mathbf{a}_{\nu }+\mathbf{a}_{\nu }\mathbf{a}_{\mu }=2\eta
_{\mu \nu },
\end{equation}%
specifically 
\end{subequations}
\begin{equation}
\mathbf{a}_{0}\cdot \mathbf{a}_{0}=-1,\text{ \ \ \ \ \ \ \ \ }dt\cdot dt=-1
\end{equation}%
and 
\begin{equation}
\mathbf{a}_{i}\cdot \mathbf{a}_{i}=1,\text{ \ \ \ \ \ \ \ \ \ \ }dx^{i}\cdot
dx^{i}=1.
\end{equation}%
We reached expressions of the type $\mathbf{\varepsilon }^{\pm }\mathbf{I}%
_{ij}^{\pm }\mathbf{P}_{l}^{\pm }$, where 
\begin{equation}
\mathbf{I}_{ij}^{\pm }=\frac{1}{2}(1\pm \mathbf{w}_{ij}),
\end{equation}%
with%
\begin{equation}
\mathbf{w}_{ij}=w_{k}\mathbf{a}_{i}\mathbf{a}_{j},\text{ \ \ \ \ \ \ \ \ \ \ 
}w_{k}\equiv dx^{ij}\equiv dx^{i}dx^{j},
\end{equation}%
and 
\begin{equation}
\mathbf{P}_{i}^{\pm }=\frac{1}{2}(1\pm dx^{i}\mathbf{a}_{i})\text{ \ \ \ \ \
\ \ \ \ \ \ \ (no sum, \ \ }i=1,2,3\text{)}
\end{equation}%
and 
\begin{equation}
\mathbf{\varepsilon }^{\pm }=\frac{1}{2}(1\mp dt\mathbf{a}_{0}).
\end{equation}%
Notice the absence of the unit imaginary, its role being played on a case by
case basis by different elements of square is minus one in the tangent
Clifford algebra defined by the first equations (2) and (3).

In the product structure, there is a\ subset of what we called mirror
elements \cite{V55}. They play a privileged role. They constitute a
commutative algebra under correlated products in the individual factor
algebras that make that product structure \cite{V51}, \cite{V55}, \cite{V56}.

For a short history of products of structures similar to those, we go as far
back as \'{E}. Cartan when he spoke of the curvatures of Euclidean
connections as being bivector-valued differential $2-$forms \cite{Cartan23}%
.\ In 1986, Oziewicz proposed a structure which looks like our product of
two Clifford algebras \cite{Oziewicz}. Of special importance is
Helmstetter's work on Clifford algebra, because he not only has a product of
such algebras but also a subset of privileged elements \cite{Helms}. See %
\cite{V55} for our superficial report of some of his related work.

But the total picture still is more sophisticated than what we have just
presented, owing to an argument to be summarized in the next section.
Briefly speaking, we shall extend these ideas to five dimensions but in such
a way that the usual relativistic metric is maintained, reinterpreted as a
null metric in five dimensions, and where equations like (1a) and (1b) lose
their independence from each other.

\subsubsection{Of the metric in Kaluza-Klein space}

The concept of metric has evolved with the growth of differential geometry.
Metric and distance were in essence the same differential invariant in
Riemannian geometry since each of them implies the other. This is not the
case in Finsler geometry. It suffices to refibrate a Euclidean bundle as a
Finslerian bundle, where the distance simply is $\omega ^{0}$, mod $\omega
^{m}$, with $\omega ^{\mu }=(\omega ^{0},\omega ^{m})$. There are infinite
metrics ( $\sum (\omega ^{\mu })^{2}=g_{\mu \nu }(x,u)dx^{\mu }dx^{\nu }$on
the same distance).

Let us now look at the metric from the perspective of the 5-D canonical
Kaluza-Klein (KK) space of a spacetime. In \cite{V51}, we considered the KK
space endowed with translation element $d\mathbf{\wp =}$ $\omega ^{\mu }%
\mathbf{e}_{\mu }+d\tau \mathbf{u}$, signature $(-1,1,1,1,-1)$ and $0=d%
\mathbf{\wp (\cdot ,\cdot )}d\mathbf{\wp .}$ We had%
\begin{equation}
d\mathbf{\wp (\cdot ,\cdot )}d\mathbf{\wp =}\text{ }0=\eta _{\mu \nu }\omega
^{\mu }\cdot \omega ^{\nu }-d\tau \cdot d\tau +2(\omega ^{\mu }\cdot d\tau )%
\mathbf{u\cdot e}_{\mu }.
\end{equation}%
If we set $\omega ^{\mu }\cdot d\tau =0,$ 
\begin{equation}
\omega ^{0}\cdot \omega ^{0}-\sum_{i}\omega ^{i}\cdot \omega ^{i}=d\tau
\cdot d\tau .
\end{equation}%
On curves, all differentials are multiples of just one. We define $\gamma $
by $d\tau =$ $\gamma ^{-1}dt$ and obtain%
\begin{equation}
\gamma ^{-2}dt\cdot dt=dt\cdot dt-\sum_{i}(u^{i})^{2}dt\cdot dt.
\end{equation}%
Hence $\gamma ^{-2}=1-\sum_{i}(u^{i})^{2}.$ Condition $\omega ^{\mu }\cdot
d\tau =0$, however, is too strong since it implies $0=\gamma
^{-1}dx^{i}\cdot dt$ and $0=\gamma ^{-1}dt\cdot dt$, which are obviously
wrong. The appropriate treatment of the metric in KK contest was given in
Section 5 of \cite{V51}. In that argument $\mathbf{u}^{2}=-1$, which was
denoted as $\mathbf{w}^{2}=-1.$ In this paper, we reserve the symbol $%
\mathbf{w}$ for other purposes.

\subsection{Group related issues}

\subsubsection{On role and relevance of group theory in high energy physics}

Gauge symmetry has to do with the form of the Dirac and K\"{a}hler's
equations. The wave function is a spinor in the first equation, but not in
general in the second one. As Kaehler's treatment of symmetry shows \cite%
{Kahler61}, \cite{Kahler62}, the $U(1)$ group and its Lie algebra may not be
the best representation of the dynamics of the electromagnetic interaction.
In fact, gauge symmetry is mentioned in just one line in Kaehler's most
comprehensive papers in spite of the fact that he gets algebraic
representations of electrons and positrons.

It is well known that not all $1-$dimensional groups are isomorphic, but all
1-dimensional Lie algebras are, since the Lie algebra is about transported
tangent spaces. $u(1)$ may, therefore, represent translations. The
idempotent (read ideal, read spinor) for time translations is associated
with $u(1)$, though not with $U(1)$.

As for $SU(2)$, it is intimately related to rotations, and so is therefore $%
su(2)$. Like translations, rotations are displacements.

On the other hand, a group like $SU(3)$ does not reveal a geometrical
interpretation of any type, nor does $U(1)\times SU(2)\times $ $SU(3)$. But
our algebraic representation of quarks based on tangent bundle related
geometry does. In a hypothetical future paradigm of the physics, the Lie
algebra of the group $U(1)\times SU(2)\times $ $SU(3)$ will still be present
in one way another, but it might not be at the fore of its description. A
direct product of groups should be a consequence of something more
fundamental, not an opportunistic representation of reality.

The point that group theory is a useful technique but no substitute for
physics was already made by Giorgi \cite{Giorgi}

\begin{quotation}
``I think that group theory, perhaps because it seems to give information
for free, has been more misused in contemporary particle physics than any
other branch of mathematics, except geometry. Students should learn the
difference between physics and mathematics from the start''.
\end{quotation}

In physics, a group theory argument usually is an insert into some other
theory, rather than an organic part of it. If, as we claim, KC is ``\textit{%
the calculus for physics''}, much of quantum mechanics is simply standard
mathematical argument within that calculus, following a few basic physical
assumptions. The claim is substantiated by the depth of results that K\"{a}%
hler himself obtained in using it in relativistic quantum mechanics \cite%
{Kahler61}, \cite{Kahler62}, by our derivation of the electromagnetic
Hamiltonian without resort to Foldy-Wouthuysen transformations \cite{V44}
and by what we are presently doing in connection with quark phenomenology.
Of special interest in this regard is his treatment of total angular
momentum \cite{Kahler62}, which we have adapted to our commutative algebra %
\cite{V55}, \cite{V56}. Readers are asked to compare it with, for instance,
the group theoretic treatment by Gilmore \cite{Gilmore}. To conclude, group
theory is very relevant for physics, when there is nothing better.

\subsubsection{Groups, SU(3) symmetry and our road to quarks}

In K\"{a}hler's sophisticated mathematical treatment, solutions with
symmetry of differential systems are more structured than in the physical
paradigm. K\"{a}hler gave the form of solutions with time translation and
rotational symmetry as \cite{Kahler61}%
\begin{equation}
u=p(\rho ,z,d\rho ,dz)\text{ }e^{im\phi -iEt}\text{ }\epsilon ^{\pm
}I_{xy}^{\ast },
\end{equation}%
The idempotents $\epsilon ^{\pm }$ and $I_{xy}^{\ast }$ are precursors of
the idempotents with similar notation given in subsubsection 2.1.2.

The relevance of phase factors is well known from the paradigm. With equal
or greater reason, idempotent factors share that relevance. But whereas $m$
and $E$ in $e^{im\phi -iEt}$ depend on what equation one is solving and what
energy and angular momentum is available, $\epsilon ^{\pm }I_{xy}^{\ast }$
is fixed for corresponding symmetries; one does not even need to know which
are the equations that have solutions of that form.

When the symmetries are time translation (or propertime translation, see
below) jointly with rotational symmetry, the factor $p$ in (11) depends only
on $\rho $, $z$, $d\rho $ and $dz$; the dependence on $dt$ (later $d\tau $,
where $\tau $ is proper time) and $d\phi $ (in disguised form, since $\rho
d\rho d\phi =dxdy$) is taken care of by the idempotent factors, and all the
dependence on $t$ (or $\tau $) and $\phi $ is in the phase factors. Hence a
limited form of $u(1)+su(2)$ is more present in those solutions of K\"{a}%
hler's equation with electromagnetic coupling than $u(1)$ is. This presence
is embodied in both $e^{im\phi -iEt}$ and $\epsilon ^{\pm }I_{xy}^{\ast }.$

The statements just made raise the well known issue of whether a tangent
bundle symmetry ---if this were the case with $SU(2)$--- can be put together
with an auxiliary bundle symmetry, as might be thought to be the case with $%
U(1)\times SU(2)\times $ $SU(3).$ Unlike $SU(2)$, there is not any geometry
in $SU(3)$, whether in the tangent bundle or in an auxiliary bundle. This
author submits that the solution to such quandaries lies in algebra
pertaining to solutions, not in groups pertaining to equations. Those are
all the group considerations that we need to make.

Our road to quarks takes place through a commutative algebra associated with
space-propertime ($x^{i},\tau $), not ($t,x^{i}$) \cite{V51}. Once
propertime translation and rotational symmetry have been jointly implemented
into a solution, there is not the option of implementing $z$-translation
symmetry, due to the fact that $d\tau $ and $dz$ do not commute.

In that commutative algebra, $z$-translation symmetry has priority over ``$%
\rho $-translation'' symmetry since, whereas the first one is a Euclidean
symmetry, the second one is not. Because of commutativity, there will be
room for the $z-$translation idempotent, but the corresponding ``phase
shift'' will not be adequate because the square of the geometric factor in
the exponent will not be minus one. As a consequence, quarks in this theory
are unstable solutions to the point that they die as they try to become; the
exponential factor makes quarks in the making overextend themselves through
the non-decaying (actually growing with distance, hence the problem)
exponential factor. No other mechanism is necessary. In modern physics, one
should be careful with whether we are trying to solve a physical problem or
one created by the limitations of the theory. Phenomenology tells us
nevertheless that they live enough to have a mass ascribed to them.

\subsubsection{U(1) x SU(2) and SU(3) versus geometry}

The physics of the paradigm did not know how many generations to expect. It
knows now, but not yet why there are three and only three generations and
colors, and why $e$, $\mu $ and $\tau $ are associated with the generations.
The algebra that we are advocating has answers for these questions.

In our replacement of scalar-valued with Clifford-valued K\"{a}hler algebra,
the idempotent for conservation of third component of angular momentum is
just one of the three idempotents with which $SU(2)$ contributes. And the
symmetries present in solutions of the form (11) are a truncated form of $%
U(1)\times SU(2)$. Since we might have chosen any of the three rotational
idempotents, the three generations are already there. Those idempotents
commute like everything else in our superseding, commutative algebra.

Concepts like roots and weights are based on Cartan's subalgebra spanned by
the subset of commutative group generators. But the commutative algebra to
which we have referred changes all that. Hence we must be aware of the fact
that some limitations of theory may simply be artifacts of a less than
adequate mathematical description. Consider, for instance, spin. In section
27 of \cite{Kahler62}, K\"{a}hler obtains a second spin term which may
cancel or equate the standard one, thus changing the perspective one may
have of the gyromagnetic ratio. He actually emphasizes this difference at
the end of his comprehensive 1962 paper \cite{Kahler62} This ratio does not
appear in the derivation of the Pauli term from the K\"{a}hler equation with
electromagnetic coupling \cite{V44}, and yet one obtains the right equation,
and the spectrum of the hydrogen atom is not affected at all \cite{Kahler62}%
. The hyperfine corrections to the gyromagnetic ratio should then be viewed
as limitations of theory, rather than involving the unnecessary concept of
gyromagnetic ratio.

Group theory does not provide a geometric interpretation for $SU(3)$, not
even in the auxiliary bundles which are the domain of present gauge theory.
Thus the concept of group as an organizing concept in high energy physics is
inferior ab initio to theory where there is geometric interpretation through
algebra for quarks, and where other concepts are superfluous. One should not
stop getting help from group theory whenever available, but it should not be
considered more than a succedaneum for other, more attractive options. The
true understanding of the unity of the non-gravitational interactions may
lie less in groups than in algebras and geometry.

In this paper, we might still make reference to $U(1)\times SU(2)\times
SU(3) $, but just as a succedaneum for eliciting the right response from
readers familiar with the paradigm's terminology. But, in reality, $%
U(1)\times SU(2)\times SU(3)$ is made esoteric by the naturalness of the
alternative: displacements and time translations as symmetries of solutions
whose algebraic factor are the aforementioned idempotents.

\section{Precursor Works}

\subsection{Schmeikal's quarks}

Early attempts at relating $SU(3)$ to Clifford algebra are due to Chisholm
and Farwell \cite{Farwell} and \cite{Chisholm}. But it was Schmeikal's
relating of idempotents to quarks that inspired the present author's
connecting of K\"{a}hler's solutions of exterior systems with quarks \cite%
{Schmeikal}. Schmeikal exploits his virtuosity in algebra. The present
author, on the other hand, develops further K\"{a}hler's treatment of
symmetry. One more difference is that he is restricted by commutativity of
idempotents. In his algebra, one has to look for it, which is not needed in
our commutative structure. We proceed to reproduce some of his findings.

The issue that would interest particle physicists is: show me the
representation of quarks, of operators and of their action on quarks.
Schmeikal answers as follows. Six color spinor spaces are defined by the
pairs $\{\mathbf{e}_{1},\mathbf{e}_{24}\}$\textbf{, }$\{\mathbf{e}_{1}%
\mathbf{,e}_{34}\}$\textbf{, }$\{\mathbf{e}_{2}\mathbf{,e}_{34}\}$\textbf{, }%
$\{\mathbf{e}_{2}\mathbf{,e}_{4}\mathbf{\},\{e}_{3}\mathbf{,e}_{14}\mathbf{%
\},\{e}_{3}\mathbf{,e}_{24}\}$\textbf{,} the $\mathbf{e}_{\mu \nu }$ being
the bivectors $\mathbf{e}_{\mu }\mathbf{\wedge e}_{\nu }$ formed from the
orthonormal $\{\mathbf{e}_{\mu }\}$ basis of tangent vectors. The primitive
idempotents associated with what he calls the first color space are 
\begin{equation}
f_{11,12}=\frac{1}{2}(1\mathbf{+e}_{1}\mathbf{)}\frac{1}{2}\mathbf{(}1%
\mathbf{\pm e}_{24}),\;\ \ \ \ \ \;f_{13,14}=\frac{1}{2}(1-\mathbf{e}_{1})%
\frac{1}{2}(1\pm \mathbf{e}_{24}).
\end{equation}%
Notice similarities and differences with K\"{a}hler's primitive idempotents %
\cite{Kahler61}. Primitive idempotents for the other color spinor spaces are
similarly constructed

Schmeikal gives his version of the standard strong interaction operators: 
\begin{subequations}
\begin{equation}
t_{z}=\frac{1}{4}(\boldsymbol{e}_{24}-\boldsymbol{e}_{124})
\end{equation}%
\begin{equation}
y=\frac{1}{6}(-2\boldsymbol{e}_{1}+\boldsymbol{e}_{24}+\boldsymbol{e}_{124})
\end{equation}%
\begin{equation}
q=\frac{1}{6}(-\boldsymbol{e}_{1}+2\boldsymbol{e}_{24}-\boldsymbol{e}_{124})
\end{equation}%
\begin{equation}
s=-f_{12}
\end{equation}%
\begin{equation}
b=\rho -s=\frac{1}{12}(3-\boldsymbol{e}_{1}-\boldsymbol{e}_{24}-%
\boldsymbol{e}_{124}),
\end{equation}%
where $\mathbf{e}_{klm}$ is a trivector and where $s$ and $b$ stand for
strangeness and baryon number. He goes on to provide the following scheme 
\begin{table}[h]
\caption{Schmeikal's Scheme for Quarks}\centering                            
\begin{tabular}{|c|c|c|c|c|}
\hline
& $f_{11}$ & $f_{12}$ & $f_{13}$ & $f_{14}$ \\ 
& $|v>$ & $|s>$ & $|u>$ & $|d>$ \\ \hline
$t_{z}$ & $0$ & $0$ & $\frac{1}{2}$ & $-\frac{1}{2}$ \\ \hline
$y$ & $0$ & $-\frac{2}{3}$ & $\frac{1}{3}$ & $\frac{1}{3}$ \\ \hline
$q$ & $0$ & $-\frac{1}{3}$ & $\frac{2}{3}$ & $-\frac{1}{3}$ \\ \hline
$s$ & $0$ & $-1$ & $0$ & $0$ \\ \hline
$b$ & $0$ & $\frac{1}{3}$ & $\frac{1}{3}$ & $\frac{1}{3}$ \\ \hline
\end{tabular}%
\end{table}
where $|v>$ stands for neutrinos and where $|s>,$ $|u>$ and $|d>$ stand for
quarks.

To conclude, Chrisholm, Farwell, Schmeikal and this author believe in our
own ways that the auxiliary bundles and internal spaces are unnecessary
concepts.

\subsection{Clifford-valued Solutions Endowed with Symmetry}

Our approach, like K\"{a}hler's, is based on solutions with symmetry of
exterior systems \cite{Kahler61}. But we differ from him in that our
idempotents are Clifford-valued \cite{V51} and that instead of considering
just binary idempotents ---i.e. products of just two monary ones as is \cite%
{Kahler61}, \cite{Kahler62}--- we consider ternary idempotents \cite{V51}, %
\cite{V55}, \cite{V56}.

In solutions with symmetry like (11), the idempotents are \textit{rigid},
unlike the exponentials, which are \textit{flexible, }meaning that there is
room in principle for different values of the coefficients in the exponents.
Although it is not a most pertinent example here for not being quantum
mechanical, the standard computation of the Compton effect illustrates that
the energies and momenta of the particles emerging from the collision take
whatever values are determined by the conservation laws and the input
values. Flexibility does not mean irrelevance. Exponential factors may still
forbid a reaction. Such is the case with antiproton decay, allowed by
crossing of neutron decay, but energetically forbidden. But the nature of
the particles in the output of a HEP collision is primarily determined by
the idempotents, whose rigidity determines what final products can emerge
from the collisions. So, the algebra of idempotents should constitute the
first approximation in the study of particle phenomenology

The third factor in (11), can be any function of the type $p(\rho ,d\rho
,z,dz)$ needed in each case to make a solution of whatever equation it
pertains.

In view of those considerations as to what is fixed in solutions with
symmetry, we proceed to discuss idempotents, for they constitute their least
flexible factor. We shall have all three options for $\mathbf{I}_{ij}^{\pm }$
and all three options for $\mathbf{P}_{m}^{\pm }$, in any combination$.$ But
we shall have no more than one $\mathbf{I}^{\pm }$ factor and not more than
one $\mathbf{P}^{\pm }$ factor in each product, since anything else would
not make sense. All these idempotents will have to be accompanied by
corresponding geometric phase factors, i.e. 
\end{subequations}
\begin{equation}
e^{m\phi \mathbf{a}_{i}\mathbf{a}_{j}}\mathbf{I}_{ij}^{\pm },\text{ \ \ \ \
\ }e^{-E\tau \mathbf{a}_{4}}\mathbf{\varepsilon }^{\pm }\text{\ \ \ \ \ \ }%
e^{\lambda _{i}x^{i}\mathbf{a}_{i}}\mathbf{P}_{i}^{\pm },
\end{equation}%
with no sums over repeated indices. The symbol $\mathbf{u}$ is more
representative than $\mathbf{a}_{4}$ \cite{V51}, but we can ignore that for
the moment.

It should be noticed that the exponentials for space-translation symmetries,
unlike the one for time-translation symmetry, are not phase factors by
virtue of the fact that the square of $\mathbf{a}_{i}$ is not minus one.
Time translation should thus have pre-eminence over the other translation
symmetries. Hence, the difference in character between the exponential
factors for space and propertime translations may be the reason why quarks
do not behave like normal particles.

In HEP phenomenology, operators do not follow the same pattern in the
electron generation as in the other generations. See, for instance, the $%
I_{3}$ of the Gellmann-Nishijima equation, where, in addition, there are too
many operators. To make matters still worse, components are not invariants.
What should matter is that hadrons should be viewed as composites with
proper value of total angular momentum (or something along those lines) and
not that individual quarks have proper valued different from zero for one of
many operators, and zero for all other operators.

\section{A Provisional Palette of Algebraic Quarks}

\subsection{Source palette of ternary idempotents}

In paper \cite{V56}, we derived the following table:

\begin{center}
Table 2. Constituent idempotents of type $\boldsymbol{\epsilon}\boldsymbol{I}%
_{12}\boldsymbol{P}$ \\[0pt]

\begin{tabular}{|c|c|c|c|}
\hline
$u/d$ & Subscript 1 & Subscript 2 & Subscript 3 \\ 
$u^{3}$ & $\boldsymbol{\epsilon}^{+}\boldsymbol{I}_{12}^{+}\boldsymbol{P}%
_{1}^{+}$ & $\boldsymbol{\epsilon}^{+}\boldsymbol{I}_{12}^{+}\boldsymbol{P}%
_{1}^{-}$ & $-\boldsymbol{\epsilon}^{+}\boldsymbol{I}_{12}^{\prime \,+}$ \\ 
\hline
$d^{3}$ & $\boldsymbol{\epsilon}^{+}\boldsymbol{I}_{12}^{-}\boldsymbol{P}%
_{2}^{+}$ & $\boldsymbol{\epsilon}^{+}\boldsymbol{I}_{12}^{-}\boldsymbol{P}%
_{2}^{-}$ & $-\boldsymbol{\epsilon}^{+}\boldsymbol{I}_{12}^{\prime \,-}$ \\ 
\hline
$\overline{d}^{3}$ & $\boldsymbol{\epsilon}^{-}\boldsymbol{I}_{12}^{+}%
\boldsymbol{P}_{2}^{-}$ & $\boldsymbol{\epsilon}^{+}\boldsymbol{I}_{12}^{+}%
\boldsymbol{P}_{2}^{+}$ & $-\boldsymbol{\epsilon}^{-}\boldsymbol{I}%
_{12}^{\prime \,+}$ \\ \hline
$\overline{u}^{3}$ & $\boldsymbol{\epsilon}^{-}\boldsymbol{I}_{12}^{-}%
\boldsymbol{P}_{1}^{-}$ & $\boldsymbol{\epsilon}^{-}\boldsymbol{I}_{12}^{-}%
\boldsymbol{P}_{1}^{+}$ & $-\boldsymbol{\epsilon}^{-}\boldsymbol{I}%
_{12}^{\prime \,-}$ \\ \hline
\end{tabular}

\bigskip
\end{center}

The superscript 3 in the first column correlates with the absence of the
subscript 3 in the other three columns. This table is for the generation of
the electron, proton and neutron, precisely because of that superscript. In
historical retrospect, one should have used first component rather than
third component of angular momentum in dealing with the electromagnetic
interaction. In that way that generation would have been associated with $%
\boldsymbol{I}_{23}$ and the superscript $1$ of $u$, rather than with $%
\boldsymbol{I}_{12}^{+}$ and the superscript 3.

We also indicated the obvious way to make parallel tables for superscripts $%
1 $ and $2$ of $u$ and $d.$ We now proceed to put all those tables together
and make the idempotents correspond to actual quarks, but only as an
illustrative example of the methodology. The palette of quarks of Table 3
follows.\bigskip

\begin{center}
Table 3. Provisional Palette of Quarks\bigskip

\begin{tabular}{|c|c|c|c|}
\hline
& Color 1 & Color 2 & Color 3 \\ \hline
$t=u^{1}$ & $-\boldsymbol{\epsilon}^{+}\boldsymbol{I}_{23}^{\prime \,+}$ & $%
\boldsymbol{\epsilon}^{+}\boldsymbol{I}_{23}^{+}\boldsymbol{P}_{2}^{+}$ & $%
\boldsymbol{\epsilon}^{+}\boldsymbol{I}_{23}^{+}\boldsymbol{P}_{2}^{-}$ \\ 
\hline
$b=d^{1}$ & $-\boldsymbol{\epsilon}^{+}\boldsymbol{I}_{23}^{\prime \,-}$ & $%
\boldsymbol{\epsilon}^{+}\boldsymbol{I}_{23}^{-}\boldsymbol{P}_{3}^{+}$ & $%
\boldsymbol{\epsilon}^{+}\boldsymbol{I}_{23}^{-}\boldsymbol{P}_{3}^{-}$ \\ 
\hline
$\bar{b}=\overline{d}^{1}$ & $-\boldsymbol{\epsilon}^{-}\boldsymbol{I}%
_{23}^{\prime \,+}$ & $\boldsymbol{\epsilon}^{-}\boldsymbol{I}_{23}^{+}%
\boldsymbol{P}_{3}^{-}$ & $\boldsymbol{\epsilon}^{-}\boldsymbol{I}_{23}^{+}%
\boldsymbol{P}_{3}^{+}$ \\ \hline
$\bar{t}=\bar{u}^{1}$ & $-\boldsymbol{\epsilon}^{-}\boldsymbol{I}%
_{23}^{\prime \,-}$ & $\boldsymbol{\epsilon}^{-}\boldsymbol{I}_{23}^{-}%
\boldsymbol{P}_{2}^{-}$ & $\boldsymbol{\epsilon}^{-}\boldsymbol{I}_{23}^{-}%
\boldsymbol{P}_{2}^{+}$ \\ \hline
$c=u^{2}$ & $\boldsymbol{\epsilon}^{+}\boldsymbol{I}_{31}^{+}\boldsymbol{P}%
_{3}^{-}$ & $-\boldsymbol{\epsilon}^{+}\boldsymbol{I}_{31}^{\prime \,+}$ & $%
\boldsymbol{\epsilon}^{+}\boldsymbol{I}_{31}^{+}\boldsymbol{P}_{3}^{+}$ \\ 
\hline
$s=d^{2}$ & $\boldsymbol{\epsilon}^{+}\boldsymbol{I}_{31}^{-}\boldsymbol{P}%
_{1}^{-}$ & $-\boldsymbol{\epsilon}^{+}\boldsymbol{I}_{31}^{\prime \,-}$ & $%
\boldsymbol{\epsilon}^{+}\boldsymbol{I}_{31}^{-}\boldsymbol{P}_{1}^{+}$ \\ 
\hline
$\bar{s}=\overline{d}^{2}$ & $\boldsymbol{\epsilon}^{-}\boldsymbol{I}%
_{31}^{+}\boldsymbol{P}_{1}^{+}$ & $-\boldsymbol{\epsilon}^{-}\boldsymbol{I}%
_{31}^{\prime \,+}$ & $\boldsymbol{\epsilon}^{-}\boldsymbol{I}_{31}^{+}%
\boldsymbol{P}_{1}^{-}$ \\ \hline
$\bar{c}=\bar{u}^{2}$ & $\boldsymbol{\epsilon}^{-}\boldsymbol{I}_{31}^{-}%
\boldsymbol{P}_{3}^{+}$ & $-\boldsymbol{\epsilon}^{-}\boldsymbol{I}%
_{31}^{\prime \,-}$ & $\boldsymbol{\epsilon}^{-}\boldsymbol{I}_{31}^{-}%
\boldsymbol{P}_{3}^{-}$ \\ \hline
$u=u^{3}$ & $\boldsymbol{\epsilon}^{+}\boldsymbol{I}_{12}^{+}\boldsymbol{P}%
_{1}^{+}$ & $\boldsymbol{\epsilon}^{+}\boldsymbol{I}_{12}^{+}\boldsymbol{P}%
_{1}^{-}$ & $-\boldsymbol{\epsilon}^{+}\boldsymbol{I}_{12}^{\prime \,+}$ \\ 
\hline
$d=d^{3}$ & $\boldsymbol{\epsilon}^{+}\boldsymbol{I}_{12}^{-}\boldsymbol{P}%
_{2}^{+}$ & $\boldsymbol{\epsilon}^{+}\boldsymbol{I}_{12}^{-}\boldsymbol{P}%
_{2}^{-}$ & $-\boldsymbol{\epsilon}^{+}\boldsymbol{I}_{12}^{\prime \,-}$ \\ 
\hline
$\overline{d}=\overline{d}^{3}$ & $\boldsymbol{\epsilon}^{-}\boldsymbol{I}%
_{12}^{+}\boldsymbol{P}_{2}^{-}$ & $\boldsymbol{\epsilon}^{-}\boldsymbol{I}%
_{12}^{+}\boldsymbol{P}_{2}^{+}$ & $-\boldsymbol{\epsilon}^{-}\boldsymbol{I}%
_{12}^{\prime \,+}$ \\ \hline
$\overline{u}=\overline{\bar{u}}^{3}$ & $\boldsymbol{\epsilon}^{-}%
\boldsymbol{I}_{12}^{-}\boldsymbol{P}_{1}^{-}$ & $\boldsymbol{\epsilon}^{-}%
\boldsymbol{I}_{12}^{-}\boldsymbol{P}_{1}^{+}$ & $-\boldsymbol{\epsilon}^{-}%
\boldsymbol{I}_{12}^{\prime \,-}$ \\ \hline
\end{tabular}%
\bigskip
\end{center}

The algebraic interpretation of quarks as idempotents speaks clearly of why
there are three and only three generations, and two families per generation.
This might be thought to be a peculiarity of the solution that we are using,
but we shall see in section (6) that the alternative solution fits a similar
pattern. The relation of the number of colors to the dimensionality of the
3-D configuration space is less obvious. We shall make clear in the next
subsection that we can make $\mathbf{P}_{1}$, $\mathbf{P}_{2}$ and $\mathbf{P%
}_{3}$ appear explicitly in columns for colors $1$, $2$ and $3$ respectively.

\subsection{Algebraic color}

In table 2, the subscript of $\mathbf{P}$ does not always correspond to the
heading of the column where the idempotents are placed. Yet color still is
closely related to that subscript, as we now explain.

In each generation use the term primary color to refer to the one (among
colors $k=1,2,3)$ which is different from the subscripts $i$ and $j$ in $%
\mathbf{I}_{ij}$. It is thus $1$, $2$ and $3$ for the respective generations
of $(t,b)$, $(c,s)$ and $(u,d).$ The other two colors will be called
secondary. The following equalities apply to the secondary colors of the
different generations:%
\begin{equation}
\mathbf{\varepsilon }^{\pm }\mathbf{I}_{ij}^{+}\mathbf{P}_{i}^{\pm }=\mathbf{%
\varepsilon }^{\pm }\mathbf{I}_{ij}^{+}\mathbf{P}_{j}^{\pm }\text{, \ \ \ \
\ \ \ }\mathbf{\varepsilon }^{\pm }\mathbf{I}_{ij}^{-}\mathbf{P}_{i}^{\pm }=%
\mathbf{\varepsilon }^{\pm }\mathbf{I}_{ij}^{-}\mathbf{P}_{j}^{\mp }.
\end{equation}%
Notice the inversion of superscripts of $\mathbf{P}$ that accompanies the
change in its subscript in the second of those equations.

The expression in this table of hypothetical algebraic quarks of primary
color resulted from the sum $\mathbf{\varepsilon }^{\pm }\mathbf{I}%
_{ij}^{\pm }\mathbf{P}_{k}^{+}+\mathbf{\varepsilon }^{\pm }\mathbf{I}%
_{ij}^{\pm }\mathbf{P}_{k}^{-}.$ We, however, wrote $\mathbf{\varepsilon }%
^{\pm }\mathbf{I}_{ij}^{\prime +}$ and $\mathbf{\varepsilon }^{\pm }\mathbf{I%
}_{ij}^{\prime -}$instead of $\mathbf{\varepsilon }^{\pm }\mathbf{I}%
_{ij}^{+} $ and $\mathbf{\varepsilon }^{\pm }\mathbf{I}_{ij}^{-}$ in order
to signify that, in principle, $\mathbf{\varepsilon }^{\pm }\mathbf{I}%
_{ij}^{\pm }\mathbf{P}_{k}^{+}$ and $\mathbf{\varepsilon }^{\pm }\mathbf{I}%
_{ij}^{\pm }\mathbf{P}_{k}^{-}$ will multiply different factors and,
therefore, the sum is not justified except as an approximation which might
work in certain circumstances.

In view of these considerations, the row for, say, quark $c$ can now be
written%
\begin{equation}
\boldsymbol{\epsilon}^{+}\boldsymbol{I}_{31}^{+}\boldsymbol{P}_{1}^{-},\ \ \
\ -(\boldsymbol{\epsilon}^{+}\boldsymbol{I}_{31}^{+}\mathbf{P}_{2}^{+}\oplus %
\boldsymbol{\epsilon}^{+}\boldsymbol{I}_{31}^{+}\mathbf{P}_{2}^{-}),\ \ \ \
\ \boldsymbol{\epsilon}^{+}\boldsymbol{I}_{31}^{+}\boldsymbol{P}_{3}^{+},
\end{equation}%
the symbol $\oplus $ being chosen simply to remind ourselves of the remark
just made about the lack of justification of adding idempotents
indiscriminately. Of course, more serious is the remark already made that
this is not at all the solution that we should be considering, were it not
for the specific purpose of demonstrating the methodology regardless of
specific claim as to the solution to which quarks would correspond.

One would assume that the notation for secondary colors illustrated in (16)
is better than the one in table 3. But this may not be so, as we now explain.

We compare%
\begin{equation*}
\begin{array}{ccc}
u_{1,2} & 1 & 2 \\ 
u & \boldsymbol{\epsilon}^{+}\boldsymbol{I}_{12}^{+}\boldsymbol{P}_{1}^{+} & %
\boldsymbol{\epsilon}^{+}\boldsymbol{I}_{12}^{+}\boldsymbol{P}_{1}^{-} \\ 
\bar{u} & \boldsymbol{\epsilon}^{-}\boldsymbol{I}_{12}^{-}\boldsymbol{P}%
_{1}^{-} & \boldsymbol{\epsilon}^{-}\boldsymbol{I}_{12}^{-}\boldsymbol{P}%
_{1}^{+}%
\end{array}%
\end{equation*}%
with%
\begin{equation*}
\begin{array}{ccc}
u_{1,2} & 1 & 2 \\ 
u & \boldsymbol{\epsilon}^{+}\boldsymbol{I}_{12}^{+}\boldsymbol{P}_{1}^{+} & %
\boldsymbol{\epsilon}^{+}\boldsymbol{I}_{12}^{+}\boldsymbol{P}_{2}^{-} \\ 
\bar{u} & \boldsymbol{\epsilon}^{-}\boldsymbol{I}_{12}^{-}\boldsymbol{P}%
_{1}^{-} & \boldsymbol{\epsilon}^{-}\boldsymbol{I}_{12}^{-}\boldsymbol{P}%
_{2}^{+}%
\end{array}%
\end{equation*}%
Because of (15), the two $u$ rows coincide, but the $\bar{u}$ rows do not.
The question then arises of which of these two equivalent representations of 
$u_{1,2}$ should we go by for the purpose of obtaining antiparticles, sure
enough through reversion of the superscripts. In the second representation $%
\bar{u}_{1}$ and $\bar{u}_{2}$ come out equal to each other by virtue of
(15)). We thus disregard this option.

The sum $u_{1}+u_{2}+u_{3}$ yields zero, ignoring again exponential factors.
But $u_{1}+u_{2}+d_{3}$ yields $\boldsymbol{\epsilon}^{+}d\mathbf{x}%
^{^{\prime }12}$ under the same circumstances. This would lead us to
consider proper values of total operators for $u_{1}+u_{2}+d_{3}$ angular
momenta, if we had to consider this palette seriously.

\subsection{Algebraic flavor}

The charge of quarks and their composites is governed by the
GellMann-Nishijima formula. But this formula should be considered as having
limitations typical of phenomenological formulas. One such limitation is the
profusion of concepts, like the potentially unnecessary large number of
operators. In order to establish notation, we write the formula as 
\begin{equation}
Q=\frac{B}{2}+Y=\frac{B}{2}+I_{3}+\frac{C+S+T+B^{\prime }}{2},
\end{equation}%
the meaning of each symbol then being easily inferred.

The asymmetry between generations is obvious in this formula. In order to
close the gap between phenomenology and algebra, we replace the operator $%
I_{3}$ of the paradigm with $\mathbf{I}_{3}$ defined as%
\begin{equation}
\mathbf{I}_{3}=\frac{\mathbf{I}_{xy}^{+}-\mathbf{I}_{xy}^{-}}{2}
\end{equation}%
If we define operators $U$ and $D$ by 
\begin{equation}
U\equiv \mathbf{I}_{xy}^{+},\;\;\;D\equiv -\mathbf{I}_{xy}^{-},
\end{equation}%
we then have 
\begin{equation}
\frac{U+D}{2}=\frac{d\mathbf{x}d\mathbf{y}}{2}=\mathbf{I}_{3}.
\end{equation}%
and momentarily view the GellMann-Nishijima equation as%
\begin{equation}
Q=\frac{B}{2}+\frac{U+D+C+S+T+B^{\prime }}{2}.
\end{equation}%
The proper values of $u$ under the action of $U$ and $D$ are $1$ and $0$,
and those for $d$ are $0$ and $-1.$

Now that all flavors are in the same footing, we proceed to view $C$, $S$, $%
T $ and $B^{\prime }$ algebraically:%
\begin{equation}
C\equiv \mathbf{I}_{yz}^{+},\;\;\;S\equiv -\mathbf{I}_{yz}^{-},\text{ \ \ }%
T\equiv \mathbf{I}_{yz}^{+},\;\;\;B\equiv -\mathbf{I}_{yz}^{-}.
\end{equation}%
and, therefore,%
\begin{equation}
\frac{C+S}{2}\equiv \frac{\mathbf{I}_{yz}^{+}-\mathbf{I}_{yz}^{-}}{2}=\frac{d%
\mathbf{y}d\mathbf{z}}{2}\mathbf{=I}_{1},\text{ \ \ \ \ }\frac{T+B^{\prime }%
}{2}\equiv \frac{\mathbf{I}_{zx}^{+}-\mathbf{I}_{zx}^{-}}{2}=\frac{d\mathbf{z%
}d\mathbf{x}}{2}\mathbf{=I}_{2}.
\end{equation}%
When dealing with ternary idempotents, we shall refer to the $\mathbf{I}%
_{ij}^{\pm }$ as isospin idempotents, and the $d\mathbf{x}^{ij}$ as isospin
operators. It is certainly the case that $u_{3}$ and $d_{3}$ are proper
functions of $d\mathbf{x}^{12}$ with proper values $\pm 1$. But the quarks
of secondary colors in the nucleons generation are not proper functions of
this operator.

In HEP\ phenomenology, operators are created and proper values are assigned
ad hoc. The specific proper values for so many operators --- $B$, $U$, $D$, $%
C$, $S$, $T$ and $B^{\prime }$--- makes it unlikely that an algebraic
representation for them exists. Replacement of equations (19) and (22) in
(21) is not meaningful. Hence, in this paper, we shall content ourselves
with a hybrid treatment lying between phenomenology and what should be an
algebraic representation. We thus build the algebraic table for the
generation of electrons and nucleons, the operators acting only on the
primary color. With $B$ as $-d\mathbf{\tau /}3$, we have the following
scheme of proper values:\bigskip

\begin{center}
Table 4. Proper Values of Quarks

$%
\begin{array}{ccccccc}
&  & -d\mathbf{\tau /}3 & d\mathbf{x}d\mathbf{y} & B/2 & I_{3} & B/2+I_{3}
\\ 
u_{3} & \mathbf{\varepsilon }^{+}\mathbf{I}_{12}^{\prime +} & 1/3 & 1 & 1/6
& 1/2 & 2/3 \\ 
d_{3} & \mathbf{\varepsilon }^{+}\mathbf{I}_{12}^{\prime -} & 1/3 & -1 & 1/6
& -1/2 & -1/3 \\ 
\bar{d}_{3} & \mathbf{\varepsilon }^{-}\mathbf{I}_{12}^{\prime +} & -1/3 & 1
& -1/6 & 1/2 & 1/3 \\ 
\bar{u}_{3} & \mathbf{\varepsilon }^{-}\mathbf{I}_{12}^{\prime -} & -1/3 & -1
& -1/6 & -1/2 & -2/3%
\end{array}%
$
\end{center}

Recall that the subscripts of $u$ and $d$ are for color. Their superscripts
have been omitted. For other generations we would replace $d\mathbf{x}d%
\mathbf{y}$ by cyclic permutation. We have connected with the phenomenology
of the paradigm, but much remains to be done.

In the second column, we left out the factors $\mathbf{P}$ because they do
not influence the action of those operators. So, the different colors are
not represented. There is nothing wrong with that. But, since the $\mathbf{%
\varepsilon }^{\pm }\mathbf{I}_{12}^{\prime \ast }$ by themselves represent
leptons and the charges than one gets are not those of leptons, one has to
conclude that the GellMann-Nishijima equation does not represent any deep
truth of nature but a construction that puts together ad hoc operators.

\section{Stoichiometry}

Stoichiometry may be used to explain reactions or, with the help of
phenomenology, find algebraic representations of different types of
particles. There is a very large number of simple reactions that may help
determine the formula for all particles. We start with a few ones. Readers
need to be aware that there are steps where a little bit of guess work is
involved, specially but not only because dealing only with idempotents is
only a first step towards a more thorough treatment. We are thus
illustrating possibilities.

We start with neutron decay. It is interpreted as a $d$ going into a $u$.
One wonders whether the other $d$ and the $u$ of the neutron just stand by
while the decay happens. When we let color enter the representation of the
decay, a more sophisticated process takes place, as we now show.

\subsection{Decay of $d$ quarks of secondary color}

Since the odd man out in the proton and neutron respectively are the $d$ and
the $u$ quarks, we reserve for these quarks the subscript $3$, i.e. the
label of their generation. Thus, a neutron will be given by ($%
d_{1},d_{2},u_{3}$) and the proton by ($u_{1},u_{2},d_{3}$). The decay of
just a $d$ into a $u$ does not explain neutron decay, for we would then
have, say, ($u_{1},d_{2},u_{3}$) instead of ($u_{1},u_{2},d_{3}$). That is
simply ad hoc interpretation when one does not take color into account.

We proceed by reverse engineering to address the issue of a $d$ quark going
to a $u$ quark$.$ We start with $u_{1}$ from which we want to reach some $d.$
A first step is motivated by the fact that neutron decay generates an
electron, whose representation was amply discussed in \cite{V51}. We get:%
\begin{equation}
u_{1}\text{ }=\mathbf{\varepsilon }^{+}\mathbf{I}_{12}^{+}\mathbf{P}_{1}^{+}=%
\text{ }\mathbf{I}_{12}^{+}\mathbf{P}_{1}^{+}-\mathbf{\varepsilon }^{-}%
\mathbf{I}_{12}^{+}\mathbf{P}_{1}^{+}=\text{ }\mathbf{I}_{12}^{+}\mathbf{P}%
_{1}^{+}\text{ }-\mathbf{\varepsilon }^{-}\mathbf{I}_{12}^{+}\text{ }+\text{ 
}\mathbf{\varepsilon }^{-}\mathbf{I}_{12}^{+}\mathbf{P}_{1}^{-}.
\end{equation}%
$\mathbf{\varepsilon }^{-}\mathbf{I}_{12}^{+}$ is an electron with
spin/chirality ``plus'', which we thus move to the left hand side to
accompany $u_{1}$:%
\begin{equation*}
u_{1}+\mathbf{\varepsilon }^{-}\mathbf{I}_{12}^{+}\text{ }=\text{ }\mathbf{I}%
_{12}^{+}\mathbf{P}_{1}^{+}+\mathbf{\varepsilon }^{-}\mathbf{I}_{12}^{+}%
\mathbf{P}_{1}^{-}\text{ }=\text{ \ \ \ \ \ \ \ \ \ \ \ \ \ \ \ \ \ \ \ }
\end{equation*}%
\begin{equation}
=\mathbf{I}_{12}^{+}\mathbf{P}_{1}^{+}+\text{ }\mathbf{I}_{12}^{+}\mathbf{P}%
_{1}^{-}-\mathbf{\varepsilon }^{+}\mathbf{I}_{12}^{+}\mathbf{P}_{1}^{-}=%
\mathbf{I}_{12}^{+}-\mathbf{\varepsilon }^{+}\mathbf{I}_{12}^{+}\mathbf{P}%
_{1}^{-}.
\end{equation}%
The last term will now be decomposed so that the minus sign at the front of
the ternary idempotent will go into a plus sign, while remaining on the same
side of the equation,%
\begin{equation}
-\mathbf{\varepsilon }^{+}\mathbf{I}_{12}^{+}\mathbf{P}_{1}^{-}\text{ }=%
\text{ }-\mathbf{\varepsilon }^{+}\mathbf{P}_{1}^{-}+\mathbf{\varepsilon }%
^{+}\mathbf{I}_{12}^{-}\mathbf{P}_{1}^{-}.
\end{equation}%
Since \textbf{$\varepsilon $}$^{+}\mathbf{I}_{12}^{-}\mathbf{P}_{1}^{-}$
equals \textbf{$\varepsilon $}$^{+}\mathbf{I}_{12}^{-}\mathbf{P}_{2}^{+}$,
we read from our palette of quarks that this is $d_{1}.$ Combining equations
(21)-(23), one gets%
\begin{equation}
d_{1}=u_{1}+\mathbf{\varepsilon }^{-}\mathbf{I}_{12}^{+}\text{ }+\text{ }%
\mathbf{\varepsilon }^{+}\mathbf{P}_{1}^{-}-\mathbf{I}_{12}^{+}.
\end{equation}%
We similarly obtain 
\begin{equation}
d_{2}=u_{2}+\mathbf{\varepsilon }^{-}\mathbf{I}_{12}^{+}\text{ }+\text{ }%
\mathbf{\varepsilon }^{+}\mathbf{P}_{1}^{+}-\mathbf{I}_{12}^{+}.
\end{equation}%
Thus the $u_{3}$ of the neutron and the $d_{3}$ of the proton should
exchange places in neutron decay. So, in this mathematically induced
HEP-like theory, the decay of the neutron is a collective phenomenon, as it
involves all three quarks of each nucleon.

\subsection{Decay of the neutron}

Addition of (24) and (25) yields%
\begin{equation}
d_{1}+d_{2}=u_{1}+u_{2}+\text{ }^{+}e\text{ }+\mathbf{\varepsilon }^{-}%
\mathbf{I}_{12}^{+}\text{ }+\text{ }\mathbf{\varepsilon }^{+}-\mathbf{I}%
_{12}^{+}-\mathbf{I}_{12}^{+},
\end{equation}%
where we have chosen to write one of the two \textbf{$\varepsilon $}$^{-}%
\mathbf{I}_{12}^{+}$ terms as $^{+}e$ to signify an electron with a defined
chirality (left superscript +). We now add $u_{3}$ to both sides of (29). We
also add zero in the form $d_{3}-d_{3}$ on the right hand side. We thus get%
\begin{equation}
d_{1}+d_{2}+u_{3}=(u_{1}+u_{2}+d_{3})+\text{ }^{+}e\text{ }+[u_{3}-d_{3}+%
\text{ }\mathbf{\varepsilon }^{+}+\text{ }\mathbf{\varepsilon }^{-}\mathbf{I}%
_{12}^{+}-\mathbf{I}_{12}^{+}-\mathbf{I}_{12}^{+}].
\end{equation}%
Let us name the six terms in the square bracket as (1) to (6) in order of
appearance. We have 
\begin{equation}
(4)+(5)=-\text{ }\mathbf{\varepsilon }^{+}\mathbf{I}_{12}^{+},\text{ \ \ \ \ 
}
\end{equation}%
\begin{equation}
(3)+(4)+(5)=\mathbf{\varepsilon }^{+}\mathbf{I}_{12}^{-},
\end{equation}%
\begin{equation}
(3)+...+(6)=\mathbf{\varepsilon }^{+}\mathbf{I}_{12}^{-}-\mathbf{I}_{12}^{+}
\end{equation}

With $\mathbf{N}$=neutron, $\mathbf{P}$=proton (We reserve $p$ for
positrons), Eq. (30) can be written as%
\begin{equation}
N=P+\text{ }^{+}e\text{ }-\text{ }\mathbf{\varepsilon }^{+}\mathbf{I}%
_{12}^{\prime +}+\mathbf{\varepsilon }^{+}\mathbf{I}_{12}^{\prime -}+\text{ }%
\mathbf{\varepsilon }^{+}\mathbf{I}_{12}^{-}\text{ }-\mathbf{I}_{12}^{+},
\label{34}
\end{equation}%
after replacing $u_{3}$ and $d_{3}$ with their algebraic expressions. Hence,
if we ignore the exponential factors hidden in $\mathbf{\varepsilon }^{+}%
\mathbf{I}_{12}^{\prime \pm }$, this yields%
\begin{equation}
\bar{\nu}_{e}=-\mathbf{\varepsilon }^{+}\mathbf{I}_{12}^{+}+\mathbf{%
\varepsilon }^{+}\mathbf{I}_{12}^{-}+\text{ }\mathbf{\varepsilon }^{+}%
\mathbf{I}_{12}^{-}\text{ }-\mathbf{I}_{12}^{+}=2\mathbf{I}_{12}^{-}\text{ }%
-(\mathbf{\varepsilon }^{+}-1)\mathbf{I}_{12}^{+}.
\end{equation}%
This is a highly dubious result. We shall not discuss why, given the caveats
already mentioned about its lack of relevance except for illustration of the
methodology.

\subsection{``Neutrinos'' and generation-independent formulas}

We now assume --just for illustrative purposes with the same generation---
that we had obtained for the quarks $u$ and $d$ of primary color the
expressions $\mathbf{\varepsilon }^{+}\mathbf{I}_{12}^{+}\mathbf{P}_{3}^{+}$
and $\mathbf{\varepsilon }^{+}\mathbf{I}_{12}^{+}\mathbf{P}_{3}^{-}$
respectively. For further illustration, we shall study all four combinations
of signs and compare results. Instead of the last four terms on the right
hand side of (29), we would have the following four combinations:%
\begin{equation}
\text{ }\mathbf{\varepsilon }^{+}\mathbf{I}_{12}^{+}\mathbf{P}_{3}^{\pm }-%
\mathbf{\varepsilon }^{+}\mathbf{I}_{12}^{-}\mathbf{P}_{3}^{\ast }+\text{ }%
\mathbf{\varepsilon }^{+}\mathbf{I}_{12}^{-}\text{ }-\mathbf{I}_{12}^{+}.
\end{equation}%
The asterisk as a superscript signifies that the sign in $\mathbf{P}%
_{3}^{\ast }$ is independent of what sign we have in the superscript of $%
\mathbf{P}_{3}$ in the first term. Let us compute for the four combinations
of signs, i.e.%
\begin{equation}
u_{3}-d_{3}=\mathbf{\varepsilon }^{+}\mathbf{I}_{12}^{+}\mathbf{P}_{3}^{\pm
}-\mathbf{\varepsilon }^{+}\mathbf{I}_{12}^{-}\mathbf{P}_{3}^{\ast }.
\label{37}
\end{equation}%
We so have the following four options%
\begin{equation}
\mathbf{\varepsilon }^{+}\mathbf{I}_{12}^{+}\mathbf{P}_{3}^{+}-\mathbf{%
\varepsilon }^{+}\mathbf{I}_{12}^{-}\mathbf{P}_{3}^{+}=\mathbf{\varepsilon }%
^{+}\mathbf{P}_{3}^{+}d\mathbf{x}d\mathbf{y.}  \label{38}
\end{equation}%
\begin{equation}
\mathbf{\varepsilon }^{+}\mathbf{I}_{12}^{+}\mathbf{P}_{3}^{-}-\mathbf{%
\varepsilon }^{+}\mathbf{I}_{12}^{-}\mathbf{P}_{3}^{-}=\mathbf{\varepsilon }%
^{+}\mathbf{P}_{3}^{-}d\mathbf{x}d\mathbf{y.}
\end{equation}%
\begin{equation}
\mathbf{\varepsilon }^{+}\mathbf{I}_{12}^{+}\mathbf{P}_{3}^{+}-\mathbf{%
\varepsilon }^{+}\mathbf{I}_{12}^{-}\mathbf{P}_{3}^{-}\text{ }=\mathbf{%
\varepsilon }^{+}\mathbf{P}_{3}^{+}(\mathbf{I}_{12}^{+}+\mathbf{I}_{12}^{-})-%
\mathbf{\varepsilon }^{+}\mathbf{I}_{12}^{-}\text{ }=\mathbf{\varepsilon }%
^{+}\mathbf{P}_{3}^{+}-\mathbf{\varepsilon }^{+}\mathbf{I}_{12}^{-}\mathbf{.}
\label{K3}
\end{equation}%
\begin{equation}
\mathbf{\varepsilon }^{+}\mathbf{I}_{12}^{+}\mathbf{P}_{3}^{-}-\mathbf{%
\varepsilon }^{+}\mathbf{I}_{12}^{-}\mathbf{P}_{3}^{+}=\mathbf{\varepsilon }%
^{+}\mathbf{I}_{12}^{+}\mathbf{P}_{3}^{-}+\mathbf{\varepsilon }^{+}\mathbf{I}%
_{12}^{-}(\mathbf{P}_{3}^{-}-1)=\mathbf{\varepsilon }^{+}\mathbf{P}_{3}^{-}-%
\mathbf{\varepsilon }^{+}\mathbf{I}_{12}^{-}.  \label{K4}
\end{equation}%
Replacement of the pair (38)-(39) in (36) yields%
\begin{equation}
\mathbf{\varepsilon }^{+}\mathbf{P}_{3}^{\pm }d\mathbf{x}d\mathbf{y+\mathbf{%
\varepsilon }^{+}\mathbf{I}_{12}^{-}-\mathbf{I}_{12}^{+}=\varepsilon }^{+}%
\frac{1}{2}(1\pm d\mathbf{x}d\mathbf{y}d\mathbf{z)-\mathbf{I}_{12}^{+},}
\end{equation}%
whose right hand side should thus be identified with $\bar{\nu}_{e}$\ on
account of (29) and well known phenomenology. If we define%
\begin{equation}
\mathbf{w}^{\pm }\equiv \frac{1}{2}(1\pm d\mathbf{x}d\mathbf{y}d\mathbf{z}),
\end{equation}%
we have%
\begin{equation}
\bar{\nu}_{e}=\mathbf{\varepsilon }^{+}\mathbf{w}^{\pm }-\mathbf{I}_{12}^{+},%
\text{ \ \ \ \ \ \ \ \ \ \ \ \ }\nu _{e}=\mathbf{\varepsilon }^{-}\mathbf{w}%
^{\mp }-\mathbf{I}_{12}^{-}.
\end{equation}%
The generation independent equation%
\begin{equation}
Z^{0}=\nu _{e}+\bar{\nu}_{e}=\mathbf{w}^{\pm }-1=-(\frac{1}{2}\mp \mathbf{w}%
),
\end{equation}%
follows. These two formulas for $Z_{0}$ correspond to%
\begin{equation}
(u_{3},\text{ }d_{3})=(\mathbf{\varepsilon }^{+}\mathbf{I}_{12}^{+}\mathbf{P}%
_{3}^{+},\text{ }\mathbf{\varepsilon }^{+}\mathbf{I}_{12}^{-}\mathbf{P}%
_{3}^{+})\text{ \ \ and \ \ }(u_{3},\text{ }d_{3})=(\mathbf{\varepsilon }^{+}%
\mathbf{I}_{12}^{+}\mathbf{P}_{3}^{-},\text{ }\mathbf{\varepsilon }^{+}%
\mathbf{I}_{12}^{-}\mathbf{P}_{3}^{-})
\end{equation}

Using now the options (37)-(38), we get, by similar process,%
\begin{equation}
\bar{\nu}_{e}=\mathbf{\varepsilon }^{+}\mathbf{P}_{3}^{\pm }-\mathbf{I}%
_{12}^{+},\text{ \ \ \ \ \ \ \ \ \ \ \ }\nu _{e}=\mathbf{\varepsilon }^{-}%
\mathbf{P}_{3}^{\mp }-\mathbf{I}_{12}^{-},
\end{equation}%
\begin{equation}
Z^{0}=\nu _{e}+\bar{\nu}_{e}=\mathbf{\varepsilon }^{+}\mathbf{P}_{3}^{\pm }+%
\mathbf{\varepsilon }^{-}\mathbf{P}_{3}^{\mp }-1=-\text{ }(\frac{1}{2}\pm d%
\mathbf{z}d\mathbf{\tau ),}
\end{equation}%
and 
\begin{equation}
(u_{3},\text{ }d_{3})=(\mathbf{\varepsilon }^{+}\mathbf{I}_{12}^{+}\mathbf{P}%
_{3}^{+},\text{ }\mathbf{\varepsilon }^{+}\mathbf{I}_{12}^{-}\mathbf{P}%
_{3}^{-})\text{ and }(u_{3},\text{ }d_{3})=(\mathbf{\varepsilon }^{+}\mathbf{%
I}_{12}^{+}\mathbf{P}_{3}^{-},\text{ }\mathbf{\varepsilon }^{+}\mathbf{I}%
_{12}^{-}\mathbf{P}_{3}^{+}).
\end{equation}%
By virtue of the fact that $Z^{0}$ can decay by similar channels of
different generations, this option would not look as attractive as the
previous one. But we still would not discard it (if this were the real
thing) since it might represent other viable channels. The phenomenology is
very rich and could help sort out these matters until the theory reaches the
point that it can stand on its own.

\subsection{Introduction to Photons}

This subsection is independent of any specific representation of quarks. The
previous manipulations may help readers feel more comfortable than if
introduced at an earlier stage.

Consider the decay of electron-positron pairs into two photons. This
corresponds to the combinations \textbf{$\varepsilon $}$^{-}\mathbf{I}%
_{12}^{+}\mathbf{+}$ \textbf{$\varepsilon $}$^{+}\mathbf{I}_{12}^{-}$ and 
\textbf{$\varepsilon $}$^{-}\mathbf{I}_{12}^{-}\mathbf{+}$ \textbf{$%
\varepsilon $}$^{+}\mathbf{I}_{12}^{+}.$ One readily gets%
\begin{equation}
\mathbf{\varepsilon }^{-}\mathbf{I}_{12}^{+}\mathbf{+\varepsilon }^{+}%
\mathbf{I}_{12}^{-}=\mathbf{\varepsilon }^{-}(\mathbf{I}_{12}^{+}-\mathbf{I}%
_{12}^{-})+\mathbf{I}_{12}^{-}=d\mathbf{x}d\mathbf{y(\varepsilon }^{-}-\frac{%
1}{2})+\frac{1}{2}=\frac{1}{2}(1+d\mathbf{\tau }d\mathbf{x}d\mathbf{y}),
\end{equation}%
\begin{equation}
\mathbf{\varepsilon }^{-}\mathbf{I}_{12}^{-}\mathbf{+\varepsilon }^{+}%
\mathbf{I}_{12}^{+}=\mathbf{\varepsilon }^{-}(\mathbf{I}_{12}^{-}-\mathbf{I}%
_{12}^{+})+\mathbf{I}_{12}^{+}=-d\mathbf{x}d\mathbf{y(\varepsilon }^{-}-%
\frac{1}{2})+\frac{1}{2}=\frac{1}{2}(1-d\mathbf{\tau }d\mathbf{x}d\mathbf{y}%
),
\end{equation}%
Solutions for the photons corresponding to (50) appear to be%
\begin{equation}
\gamma _{1}=\frac{1}{2}(1+d\mathbf{x})\frac{1}{2}(1+d\mathbf{\tau }d\mathbf{y%
}\text{), \ \ \ \ \ \ }\gamma _{2}=\frac{1}{2}(1-d\mathbf{x})\frac{1}{2}(1-d%
\mathbf{\tau }d\mathbf{y}\text{),}
\end{equation}%
and, for (51):%
\begin{equation}
\gamma _{1}=\frac{1}{2}(1+d\mathbf{x})\frac{1}{2}(1-d\mathbf{\tau }d\mathbf{y%
}\text{), \ \ \ \ \ \ }\gamma _{2}=\frac{1}{2}(1-d\mathbf{x})\frac{1}{2}(1+d%
\mathbf{\tau }d\mathbf{y}\text{).}
\end{equation}

All these formulas correspond to defined helicities. It is then clear that
linear polarizations must correspond to combinations of the two helicities.
We do not enter into the subtle issue of opposite directions of the photons.
Let us just intimate the following. These are space-propertime
configurations of the field pertaining to photons. They would be used in
reactions representative of the exchange of a photon. This is a tricky
issue. We have avoided issues like this because it involves taking risks,
whether one is right or wrong. This author believes that it is more
productive and less controversial to obtain results that are close to the
mathematics, minimizing to any possible extent controversial issues.

Let us make, however, a remark of a general nature about trajectories. A
trajectory is a curve, thus a $1-$dimensional manifold, in time-space. All
differential $1-$forms must be multiples of just one, say $dt$. The $dx^{i}$
in $d\mathbf{x}$, $d\mathbf{y}$ and $d\mathbf{z}$ become multiples of $dt$
through the natural lifting conditions $dx^{i}-u^{i}dt=0.$ But the treatment
of the relation between time-space and space-propertime equations is not
obvious when anything going at the speed of light is involved. We leave this
problem for a future paper. These two options differ by a reversal of the
sign of the superscript of the $\mathbf{P}$'s. We do not know at this point
the implications of this difference.

\section{Path to the Palette of Algebraic Quarks}

The present paper and the previous one \cite{V56} has its origin in the
study of proper values of what we have called total operators. We use this
term to refer to products of operators where one of them is total angular
momentum. Total is used here in the sense not only that spin is included,
but specially in the sense that all three components enter the operator at
the same time and on an equal footing. Total operators do not have as proper
functions the ternary idempotents that are proper functions of components of
angular momentum \cite{V56}. In that study \cite{V56}, one had to deal with
combinations of ternary idempotents, and where the equations not only had a
proper value term on the right hand side, but also a ``covalue'' term. Those
linear combinations are the sums of the lines in table 3. The terms in each
line add up to zero, but they will certainly not when multiplied by factors,
different from one term to another. Having said that, let us stay at the
level of idempotents, disregarding those factors. There is much to learn
from them. The terms that constituted those ``zero sums splits'' led us to
the considerations of the previous two sections. They are full of HEP\
flavor, in more than one sense of the term.

\subsection{Zero Proper Values for Total Operators}

Recall the equation what we had in \cite{V56}: 
\begin{equation}
\lbrack (K+1)d\boldsymbol{r}]X_{A}=\mu ^{\prime }X_{A}+\pi _{A},
\end{equation}%
where $(K+1)$ is K\"{a}hler's total angular momentum operator. The $d%
\boldsymbol{r}$ is the usual differential element of translation. We labeled
the ternary idempotents as $X_{A}$. We tried to make sums of $X_{A}$'s
belonging to the same generation and proper value $\mu ^{\prime }$, allowing
for the extra terms $\pi _{A}.$ This rather than being an equation is an
implicit definition of terms. $\pi _{A}$ will not be a number in general We
tried to make linear combinations of $X_{A}$'s such that the sum of the $\pi
_{A}$'s for suitable $\mu ^{\prime }$ would are numbers.

We defined 
\begin{equation}
\mu :=-\frac{\mu ^{\prime }}{4}
\end{equation}%
so that Eq. (54) could be given the form%
\begin{equation}
\lbrack (K+1)d\boldsymbol{r}+4\mu ]X_{A}=\pi _{A}.
\end{equation}%
The factor $-1/4$ in $(56)$ was chosen to minimize clutter.

We formed linear combinations $\Sigma _{A}\lambda _{A}X_{A}$. We then
proceeded to compute 
\begin{equation}
\lambda _{A}[(K+1)d\boldsymbol{r}+4\mu ]X_{A}
\end{equation}%
for the 8 ternary idempotents of the generation (excluding antiquarks) and
summed all of that up: 
\begin{equation}
\sum_{A}\lambda _{A}[(K+1)d\boldsymbol{r}+4\mu ]X_{A}=\sum_{A}\pi _{A}.
\end{equation}%
Hence $\sum \pi _{A}$ became the covalue of the equation 
\begin{equation}
\lbrack (K+1)d\boldsymbol{r}][\sum_{1}^{8}\lambda _{A}X_{A}]=\mu ^{\prime
}\sum \lambda _{A}\chi _{A}+\pi ,
\end{equation}%
where $\pi $ is $\sum \pi _{A}$.

We got the following equations 
\begin{equation}
(\lambda _{1}+\lambda _{2})+(\lambda _{5}+\lambda _{6})+\mu \lbrack (\lambda
_{1}-\lambda _{2})+(\lambda _{3}-\lambda _{4})]=0,  \label{60}
\end{equation}%
\begin{equation}
(\lambda _{1}+\lambda _{2})+(\lambda _{5}+\lambda _{6})+\mu \lbrack (\lambda
_{1}-\lambda _{2})-(\lambda _{3}-\lambda _{4})]=0,
\end{equation}%
or the simpler equivalent pair%
\begin{equation}
\lambda _{4}=\lambda _{3}  \label{62}
\end{equation}%
\begin{equation}
(\lambda _{1}+\lambda _{2})+(\lambda _{5}+\lambda _{6})+\mu (\lambda
_{1}-\lambda _{2})=0.
\end{equation}%
We also got the equations 
\begin{equation}
\lbrack (\lambda _{1}-\lambda _{2})+(\lambda _{3}-\lambda _{4})]+(\lambda
_{5}-\lambda _{6})=0,
\end{equation}%
and%
\begin{equation}
\frac{1}{2}[(\lambda _{1}-\lambda _{2})-(\lambda _{3}-\lambda
_{4})]+(\lambda _{5}-\lambda _{6})=0.  \label{65}
\end{equation}%
In view of (62), they both became%
\begin{equation}
(\lambda _{1}-\lambda _{2})+2(\lambda _{5}-\lambda _{6})=0.
\end{equation}%
We then obtained 
\begin{equation}
(\lambda _{1}+\lambda _{2})-(\lambda _{3}+\lambda _{4})+(\lambda
_{5}+\lambda _{6})-(\lambda _{7}+\lambda _{8})+2\mu \lbrack (\lambda
_{5}-\lambda _{6})-(\lambda _{7}-\lambda _{8})]=0,
\end{equation}%
\begin{equation}
(\lambda _{1}+\lambda _{2})+(\lambda _{3}+\lambda _{4})+(\lambda
_{5}+\lambda _{6})+(\lambda _{7}+\lambda _{8})+2\mu \lbrack (\lambda
_{5}-\lambda _{6})+(\lambda _{7}-\lambda _{8})]=0.
\end{equation}%
Adding and subtracting (67) and (68), we got 
\begin{equation}
(\lambda _{1}+\lambda _{2})+(\lambda _{5}+\lambda _{6})+2\mu (\lambda
_{5}-\lambda _{6})=0
\end{equation}%
\begin{equation}
(\lambda _{3}+\lambda _{4})+(\lambda _{7}+\lambda _{8})+2\mu (\lambda
_{7}-\lambda _{8})=0.  \label{70}
\end{equation}%
Finally, we also got 
\begin{align}
& (\lambda _{1}-\lambda _{2})+\frac{1}{2}(\lambda _{5}-\lambda _{6})-\frac{1%
}{2}(\lambda _{7}-\lambda _{8})+  \notag \\
& +\mu \lbrack (\lambda _{1}+\lambda _{2})-(\lambda _{3}+\lambda
_{4})+(\lambda _{5}+\lambda _{6})-(\lambda _{7}+\lambda _{8})]=0.
\end{align}

We proceeded to solve this system of equations. It follows from (63) and
(69) that 
\begin{equation}
\mu (\lambda _{1}-\lambda _{2})-2\mu (\lambda _{5}-\lambda _{6})=0.
\end{equation}%
Hence, either $\mu =0$ or 
\begin{equation}
\lambda _{1}-\lambda _{2}=2(\lambda _{5}-\lambda _{6}).
\end{equation}%
In \cite{V56}, we solved the option $\mu =0$, though not exhaustively. We
shall come back to this later$.$ We shall now deal with the option $\mu \neq
0$, (73).

\subsection{Non-zero proper values}

From (64) and (73), using (62), we have%
\begin{equation}
\lambda _{6}=\lambda _{5},\text{ \ \ \ \ \ \ \ \ }\lambda _{2}=\lambda _{1}.
\end{equation}%
Then, from (63),%
\begin{equation}
\lambda _{5}=-\lambda _{1}.
\end{equation}%
From (70),%
\begin{equation}
2\lambda _{3}+(\lambda _{7}+\lambda _{8})+2\mu (\lambda _{7}-\lambda _{8})=0.
\label{76}
\end{equation}%
And from (71), 
\begin{equation}
\frac{1}{2}(\lambda _{7}-\lambda _{8})+2\mu \lambda _{3}+\mu (\lambda
_{7}+\lambda _{8})]=0.  \label{77}
\end{equation}%
The last two together yield $\mu =\pm 1/2.$ We thus get%
\begin{eqnarray}
\mu &=&\frac{1}{2},\text{ \ \ \ \ \ \ \ \ \ \ \ \ \ \ \ }\lambda
_{7}=-\lambda _{3},  \notag \\
\lambda _{A} &=&\lambda _{1},\text{ }\lambda _{1},\text{ }\lambda _{3},\text{
}\lambda _{3}\text{, }-\lambda _{1},\text{ }-\lambda _{1},\text{ }-\lambda
_{3},\text{ }\lambda _{8}.
\end{eqnarray}%
Similarly%
\begin{eqnarray}
\mu &=&-\frac{1}{2},\text{ \ \ \ \ \ \ \ \ \ \ \ \ \ \ \ }\lambda
_{8}=-\lambda _{3},  \notag \\
\lambda _{A} &=&\lambda _{1},\text{ }\lambda _{1},\text{ }\lambda _{3},\text{
}\lambda _{3}\text{, }-\lambda _{1},\text{ }-\lambda _{1},\text{ }\lambda
_{7},\text{ }-\lambda _{3}.
\end{eqnarray}

Consider (78). Two obvious choices are $\lambda _{1}=0$ and $\lambda _{3}=0$
respectively. The pattern for $\mu =0$ repeats itself if we choose $\lambda
_{3}=0$, meaning that the two idempotents $X_{5}$ and $X_{6}$ combine. On
the other hand, if we chose $\lambda _{1}=0$ and also $\lambda _{8}=0$, we
are going to have a third color that will be more like in the subsection of
neutrinos, which gave rise to appealing results. We thus suggest focussing
on 
\begin{equation}
\mu =\frac{1}{2},\text{ \ \ \ \ \ \ \ \ \ \ \ \ \ \ \ }\lambda _{A}=0,\text{ 
}0,\text{ }\lambda _{3},\text{ }\lambda _{3},\text{ }0,\text{ }0,\text{ }%
-\lambda _{3},\text{ }0.
\end{equation}%
Similarly for (79),%
\begin{equation}
\mu =-\frac{1}{2},\text{ \ \ \ \ \ \ \ \ \ \ \ \ \ \ \ }\lambda _{A}=0,\text{
}0,\text{ }\lambda _{3},\text{ }\lambda _{3},\text{ }0,\text{ }0,\text{ }0,%
\text{ }-\lambda _{3}.
\end{equation}%
The idempotents $X_{3},$ $X_{4},$ $X_{7}$ and $X_{8}$ were defined as 
\begin{equation}
X_{3}=\boldsymbol{I}_{12}^{-}\boldsymbol{P}_{1}^{+},\text{ \ \ }X_{4}=%
\boldsymbol{I}_{12}^{-}\boldsymbol{P}_{1}^{-},\text{ \ \ }X_{7}=%
\boldsymbol{I}_{12}^{-}\boldsymbol{P}_{3}^{+},\text{ \ \ }X_{8}=%
\boldsymbol{I}_{12}^{-}\boldsymbol{P}_{3}^{-},
\end{equation}%
(ignoring at this point the factor $\mathbf{\varepsilon }$, since $(K+1)d%
\boldsymbol{r}$ does not act on it). We would proceed to build the palette
of quarks from these solutions as we did for $\mu =0$ \ We would then
compute $\pi $ from the last column of a table like the present Table 3 \cite%
{V56}

\section{Conclusion}

We have found that there are three proper values, 1/2, 0 and -1/2. These
options have suboptions. Some look more interesting than other, but all of
them might be relevant. The stage is now set for being more subtle with the
study of solutions. In particular, we chose $\lambda _{2}=0$ in \cite{V56}.
We would not do so now.

As we said in the abstract, we are here at a relatively easy point for entry
to this theory by HEP\ physicists, given their knowledge of the
phenomenology. Two issues, however, will require deep knowledge of the K\"{a}%
hler calculus rather than deep knowledge of the phenomenology. One of them
is the long standing issue of the spin of nucleons, starting from the spin
of the quarks. The other issue is the GellMann-Nishijima formula. We have
said enough about it in the subsection on algebraic flavor to understand, at
tleast in the present context, that there should be an equation dealing with
the same subject but with fewer operators. We also suspect that it is a
disguided form of a value and covalue equation.

\section{Acknowledgements}

Conversations with Prof. Z. Oziewicz, Prof. D. G. Torr and Dr. B. Schmeikal
are acknowledged. Funding from PST\ Associates is deeply appreciated.

\end{document}